\pdfoutput=1

\documentclass[11pt]{article}

\usepackage{times}
\usepackage{latexsym}
\usepackage{algorithm}
\usepackage{algpseudocode}
\usepackage{amsmath}
\usepackage{graphicx}
\usepackage[preprint]{acl}

\usepackage{times}
\usepackage{latexsym}

\usepackage[T1]{fontenc}

\usepackage[utf8]{inputenc}

\usepackage{microtype}

\usepackage{inconsolata}

\usepackage{graphicx}

\title{Dynamic Grid Trading Strategy: From Zero Expectation to Market
Outperformance}

\author{Kai-Yuan Chen\thanks{Equal contribution.} \\
  Department of Computer Science \\\& Information Engineering, \\National Taiwan University\\ \texttt{a1041228@gmail.com} \\
    \And
  Kai-Hsin Chen\footnotemark[1] \\  
  Department of Physics, \\  National Taiwan University\\
  \texttt{colachen1@gmail.com} \\
    \And
  Jyh-Shing Roger Jang\thanks{Corresponding author: jang@csie.ntu.edu.tw} \\ 
  Department of Computer Science \\\& Information Engineering, \\National Taiwan University\\ 
  \texttt{jang@csie.ntu.edu.tw} \\
}

\usepackage{amsmath}
\begin{document}
\maketitle
\begin{abstract}
In this paper, we propose a profitable strategy for the cryptocurrency market through grid trading. We start by analyzing the expected value of the traditional grid trading strategy. Based on our findings, we introduce an improved approach, the \textit{Dynamic Grid-based Trading (DGT)} strategy, which demonstrates superior performance compared to conventional methods.  
\end{abstract}

\section{Introduction}

Cryptocurrencies are digital assets secured by cryptography, operating on decentralized blockchain networks. In recent years, the use of cryptocurrencies for transactions has significantly increased. As of \textit{August 17, 2024}, the daily trading volume of Bitcoin had reached approximately \textit{\$37 billion}. Furthermore, compared to traditional stock markets, cryptocurrencies exhibit substantial volatility.

To capitalize on this volatility, many traders employ grid trading strategies in their transactions. We began by proposing mathematical hypotheses and conducting proofs, which informed the design of our strategy based on the \textit{traditional grid trading model}. Following this, we developed a backtesting system to assess its effectiveness through programming.

\subsection{Traditional Grid Trading Strategy}
We will provide a brief introduction to the \textit{spot grid trading strategy} applied in our research. The grid type used in this study is the \textit{geometric grid}, which places orders at price intervals that adjust proportionally. This method enables a more efficient capture of market movements across different levels of volatility.

\subsubsection{Initial Setup}
Assume there are \( n \) grids, resulting in \( n+1 \) grid levels. Given that the grid intervals have equal ratios, let the grid size be \( k \). The current grid level is represented in black, while the other grid levels are shown in gray (as \textit{Figure 1}). Starting the grid at price \( P \), if there are \( m \) gray grid levels above and \( n-m \) gray grid levels below the current price, the corresponding grid levels are outlined in \textit{Table 1} above.

\begin{table}
  \centering
  \begin{tabular}{lc}
    \hline
    \textbf{Grid Levels} & \textbf{Prices} \\
    \hline
    \verb|Top|       & $P * (1+k)^{m}$           \\
    \verb|...|       & ...           \\
    \verb|Above|     & $P * (1+k)$           \\
    \verb|Reference| & $P$          \\
    \verb|Below|     & $P *(1+k)^{-1}$           \\
    \verb|...|       & ...           \\
    \verb|Bottom|    & $P * (1+k)^{-(n-m)}$           \\\hline
  \end{tabular}
  \caption{Structure of geometric grid}
\end{table}

If we invest \textit{M USDT} in total, the instant starting the grid, we spend \textit{$M * \frac{m}{n}$ USDT} to buy cryptocurrency and leave\textit{ $M * \frac{n-m}{n}$} as \textit{USDT}.

\subsubsection{Algorithm}
After initiating the grid strategy, a transaction is triggered only when the price crosses a gray grid level. If the price rises and crosses the \( G_i^{th} \) gray grid level, counted from the upper limit, we will sell \( \frac{1}{G_i} \) of the cryptocurrency we currently hold. The original black grid level is then marked gray, and the crossed gray level is marked black. Conversely, if the price falls and crosses the \( G_j^{th} \) gray grid level, counted from the lower limit, we will spend \( \frac{1}{G_j} \) of the \textit{USDT} we hold to purchase cryptocurrency. Similarly, the original black level is marked gray, and the crossed gray level is marked black. This dynamic adjustment of grid levels ensures that the strategy continually adapts to market movements without manual intervention.

\subsubsection{Termination}
Once the price exceeds the upper limit or falls below the lower limit of the grid, the strategy terminates. According to the algorithm's logic, if the price surpasses the upper limit, we will retain the original \textit{USDT} invested, the additional \textit{USDT} earned from grid trading, and no cryptocurrency. Conversely, if the price drops below the lower limit, all the \textit{USDT} initially invested will be converted into cryptocurrency, and we will still retain the bonus \textit{USDT} earned from grid trading.

\subsection{Data Resources}
We obtained the spot data from \textit{Binance} using the API: {https://api.binance.com/api/v3/klines}. 
The data spans from \textit{January 2021 to July 2024}, with a time interval of \textit{1 minute}, representing the highest frequency candlestick data available. Given this interval, we assume that the grid strategy executes no more than one transaction per minute. Additionally, we account for transaction fees, applying a fee of \textit{0.0008}, which corresponds to the \textit{Level 1} maker fee on OKX.

\section{Observations}
In the initial phase of our research, we identified several market patterns and conditions that guided the development of a series of hypotheses and mathematical derivations. These insights inspired the design of a novel grid trading strategy.

\subsection{Claim 1: Infinite Resources}
The first claim posits that if we have infinite capital and time, the grid trading strategy is inherently profitable. With unlimited funds, the strategy can capitalize on every price movement, as orders are consistently placed at regular intervals above and below the current price. This guarantees continuous buying at low prices and selling at higher ones, profiting from each market fluctuation. Infinite capital allows the strategy to endure even the most challenging market conditions without depleting funds. Similarly, infinite time ensures the ability to wait out any downturns or periods of stagnation, eventually profiting as the price moves. Therefore, with infinite resources, the grid trading strategy would always yield positive returns over time.

\subsection{Claim 2: Zero Expected Value}
We assume that the price either increases or decreases by \( k \) with equal probability (50-50), which is a reasonable hypothesis for a basic investor model. Additionally, we do not consider transaction fees and assume that all cryptocurrency is sold if the price falls below our grid. Under these conditions, we can assert that the expected value of a grid trading strategy with grid size \( k \) and a finite number of grid levels is zero. The proof is as follows:

\subsubsection{Required Number of Arbitrage Times}
To determine when a grid trading strategy has a positive expected value, consider investing \( M \) USDT in the grid strategy with \( n \) grids, resulting in \( n+1 \) grid levels. Since this is a spot grid trading strategy, if the price rises linearly and surpasses the highest grid level, we can earn \( P_u \) USDT, where
\begin{equation}
  P_u = \frac{M}{n} \sum_{i = 1}^{\frac{n}{2}}i
\end{equation}
Conversely, if the price falls linearly and drops below the lowest grid level, we incur a loss of $L_l$ \textit{USDT}, where 
\begin{equation}
  L_l = \frac{M}{n} [\ (\frac{n}{2})^2+\sum_{i = 1}^{\frac{n}{2}-1}i\ ]
\end{equation}
Thus, without accounting for profits from arbitrage, the expected value of the grid strategy is $E(G)$ \textit{USDT}. 
\begin{equation}
  E(G) = \frac{1}{2} * (P_u - L_l) = -\frac{M}{n}(\frac{{n}^2}{8} - \frac{n}{4})
\end{equation}

\begin{figure}[h!]
    \includegraphics[width=\columnwidth]{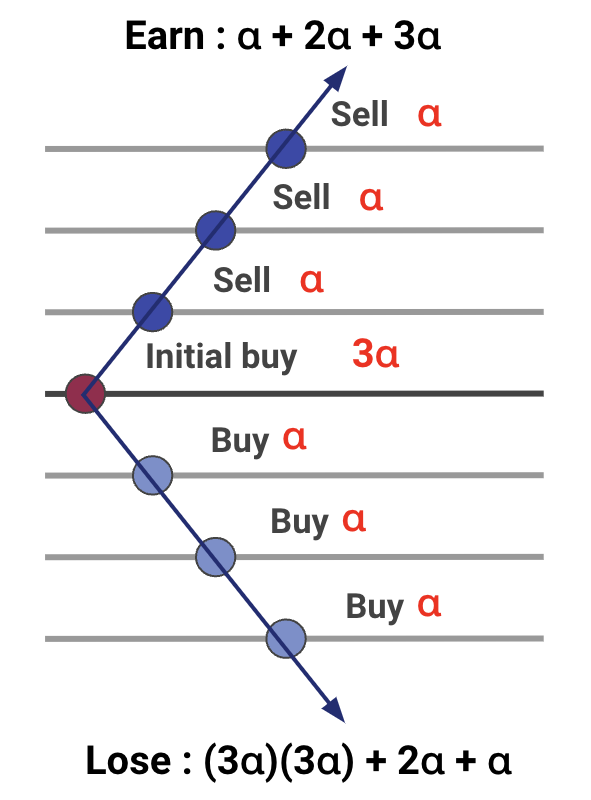}
  \caption{Grid trading detail information}
  \label{fig:grid}  
\end{figure}
In \textit{Figure 1}, we use \( n = 6 \) for demonstration. Since the grid strategy realizes profits before the price reaches the upper limit and buys more when the price falls, it results in a negative expected value if the price moves linearly. According to the derivation above, if the strategy can achieve more than \( \frac{n^2}{8} - \frac{n}{4} \) arbitrage opportunities, it will yield a positive arbitrage value.

\subsubsection{Expected Value of Grid Trading}
To calculate the expected value of transactions executed by a grid strategy, given that the probability of the price rising or falling by \( k \) is equal, we can reframe the problem as follows: \textit{What is the expected number of coin tosses required to achieve either \( \frac{n}{2} \) more heads than tails or \( \frac{n}{2} \) more tails than heads?} To address this, we model the problem using the finite automaton shown in \textit{Figure 2}. Here, \( q_i \) represents the state where the difference between the number of heads and tails is \( i \), and \( E_i \) denotes the expected value starting from state \( q_i \). The initial state is \( q_0 \), and the terminal state is \( q_{\frac{n}{2}} \). As illustrated in \textit{Figure 2}, we derive the following relationships: \( E_0 = E_1 + 1 \), \( E_1 = \frac{1}{2}(E_0 + 1) + \frac{1}{2}(E_2 + 1) \), and so forth.

\begin{figure}[h!]
    \includegraphics[width=\columnwidth]{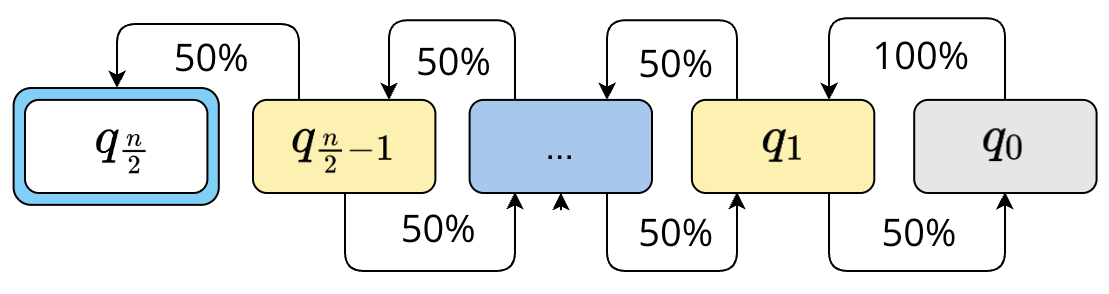}
  \caption{Finite automaton of grid trading}
  \label{fig:state}  
\end{figure}

\textbf{Proof :}

The following equations represent the expected value relationships described above:
\begin{equation}
\begin{split}
    &E_1 = E_0 - 1 \\
    &E_2 = 2E_1 - E_0 - 2 \\
    &E_3 = 2E_2 - E_1 - 2 \\
    &E_i = 2E_{i-1} - E_{i-2} - 2 \\
\end{split}
\end{equation}
From these observations, we conjecture that the recursive relationship for the expected value is \textbf{$E_m = E_0 - m^2$}. We will prove this conjecture using mathematical induction.

\subsection*{\textbf{Base Case :}} If$\ m = 1, 2$
\[
E_1 = E_0 - 1 
\]
\[
E_2 = E_0 - 4 
\]
The base case holds for $m=1, 2$.

\subsection*{Inductive Hypothesis :} 

Suppose the result holds for $m = k, k+1, k \in \mathbf{N}$

\[
E_k = E_0 - k^2 
\]

\[
E_{k+1} = E_0 - (k+1)^2
\]

\subsection*{Inductive Step}

For $m = k + 2$

\[
E_{k+2} = 2E_{k+1} - E_{k} - 2
\]

By the inductive hypothesis, we can replace the sum of the first \( k \) integers:

\[
E_{k+2} = 2[E_0 - (k+1)^2] - (E_0 - k^2) - 2
\]

Hence,

\[
E_{k+2} = E_0 - (k+2)^2
\]

\subsection*{Conclusion}
Since the base case holds and the inductive step has been proven, by mathematical induction, the statement is true for all \( n \geq 1 \).

By the proof we conducted, if the terminate state is $q_{\frac{n}{2}}$, the expected value follows the relationship given below, where $E_{\frac{n}{2}}=0$ :

\[
E_{\frac{n}{2}} = E_{0} - (\frac{n}{2})^2 
\]

\[
E_{0} = \frac{n^2}{4}
\]
Therefore, the expected value of transactions is \( \frac{n^2}{4} \). Subtracting the transaction costs due to the price difference of \( \frac{n}{2} \) and then dividing by 2—since each pair of buy and sell transactions constitutes one arbitrage opportunity—yields:
\begin{equation}
    \frac{\frac{n^2}{4}-\frac{n}{2}}{2} = \frac{n^2}{8} - \frac{n}{4}
\end{equation}
We obtain a result of \( \frac{n^2}{8} - \frac{n}{4} \), which matches the required number of arbitrage opportunities. Therefore, we can conclude that the expected value of a grid strategy is zero under our assumptions.

\begin{equation}
    E(grid\ strategy) = 0
\end{equation}

\section{Dynamic Grid-Based Trading Strategy}
Based on our observations, we have found that a grid strategy with infinite capital and time can generate a positive return, while a grid strategy that stops once the price exceeds its boundaries cannot. The key difference is that the former continues to capitalize on arbitrage opportunities, whereas the latter does not. Therefore, even though we lack infinite capital, it is unwise to stop the strategy when the price surpasses our upper or lower limits. This reasoning is straightforward, as the fundamental concept of grid trading is to buy more when the price falls. Selling all cryptocurrency when the price drops below the grid contradicts this core principle. Building on these insights, we developed a novel approach, the \textbf{\textit{Dynamic Grid-based Trading (DGT) }} strategy, which continues to operate regardless of price movements.

\subsection{DGT Algorithm}
We selected two mainstream cryptocurrencies, \textit{Bitcoin} and \textit{Ethereum}, as the assets for backtesting our strategy. As previously mentioned, we utilized \textit{1-minute candlestick data} for the testing. The algorithm for our \textbf{\textit{DGT}} strategy is outlined as follows:

The strategy resets the grid whenever the price breaks above the upper limit or falls below the lower limit, with the current price becoming the new center. If the price breaks above the upper limit, the initial capital is fully recovered and reinvested in the next grid. Conversely, if the price falls below the lower limit, the strategy holds the cryptocurrency, using the arbitrage profits gained as the principal for the new grid.

\begin{algorithm}
\caption{DGT Strategy}
\begin{algorithmic}[1]
    \State \textbf{Set Strategy Parameters:} \textit{grid sizes, levels, transaction fee}
    \State \textbf{Define Grid Calculation:} Compute grid levels around start price

    \State \textbf{Prepare Results Storage:} Initialize empty list for results

    \For{\textbf{each} combination of grid parameters}
        \State {\textbf{Initialize variables:}} \textit{wallet, input money, grid count} setup
        \For{\textbf{each} price data point}
            \If{price moves up}
                \State Execute sell if crossing grid level
            \ElsIf{price moves down}
                \State Execute buy if crossing grid level
            \EndIf
            \If{price exceeds grid limitaries}
                \State \textbf{Calculate} \textit{profit}, \textbf{Reset} \textit{grid}             \\
\ \ \ \ \ \ \ \ \ \ \ \ \ \ \ \ \ \ \textbf{Update} \textit{wallet}
            \EndIf
        \EndFor
        \State Calculate final performance metrics
    \EndFor

\end{algorithmic}
\end{algorithm}

The \textit{\textbf{wallet}} tracks the cumulative profits earned from grid arbitrage, while \textbf{\textit{input money}} represents the total invested capital. These variables are updated each time the grid is reset.

\subsection{Backtesting Results}
We used \textit{Python} to construct the entire backtesting framework. \textit{Figure 3 and 4} illustrate the performance of the \textit{\textbf{DGT}} in terms of \textit{annualized return (IRR)} across various parameter combinations for two major cryptocurrencies, \textit{BTC} and \textit{ETH}, over the backtesting period from \textit{January 2021 to July 2024}.

The \textit{\textbf{grid size}} represents the ratio of each step in the geometric grid, while \textit{\textbf{grid numbers half}} refers to the number of grids established above and below the central price when the grid is activated.

From \textit{Figure 3 and 4}, we can derive the following conclusions: First, the \textit{IRR} of our strategy remains consistently positive throughout the backtesting period, reaching as high as 60-70\%. This strong performance is largely due to the significant rise in cryptocurrency prices in recent years, which is highly favorable for spot grid trading. Moreover, \textit{ETH} demonstrates a higher \textit{IRR} than \textit{BTC}, likely because \textit{ETH}'s smaller market volume leads to greater price volatility, which benefits grid trading. Finally, a small \textit{grid size} results in lower profits due to the high proportion of transaction fees within the arbitrage profit. Conversely, a large \textit{grid size} and \textit{grid numbers half} lead to poor \textit{IRR} performance, as the infrequent activation of the grid results in lower cryptocurrency holdings. Therefore, optimizing the balance between grid size and grid numbers half is crucial for maximizing profits while minimizing costs and maintaining an active trading frequency.

\begin{figure}[t]
  \includegraphics[width=\columnwidth]{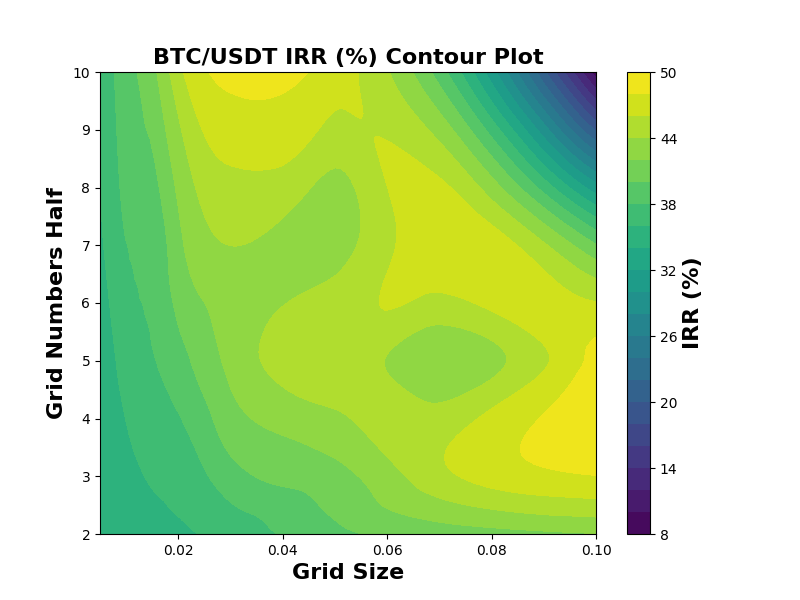}
  \caption{DGT strategy performance of \textit{BTC}}
\end{figure}

\begin{figure}[t]
  \includegraphics[width=\columnwidth]{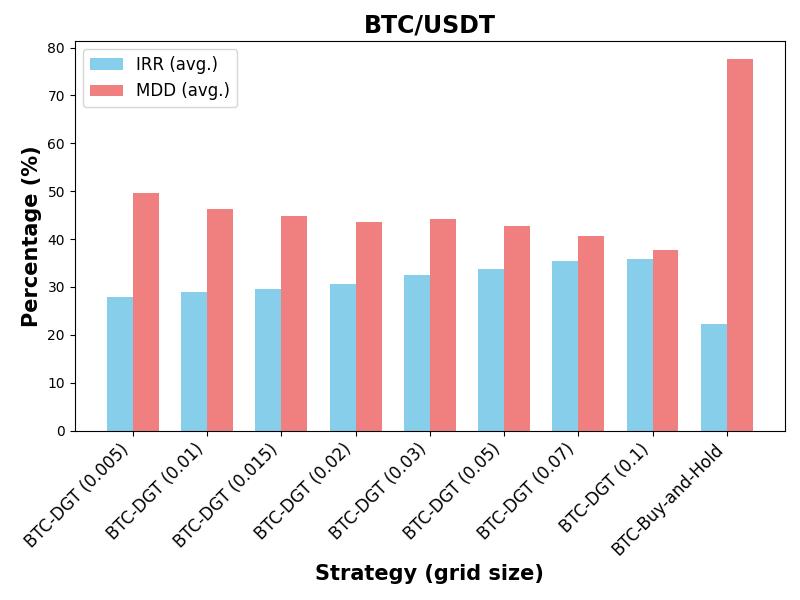}
  \caption{\textit{DGT strategy} performance compared with \textit{buy-and-hold} on BTC}
\end{figure}
\begin{figure}[t]
  \includegraphics[width=\columnwidth]{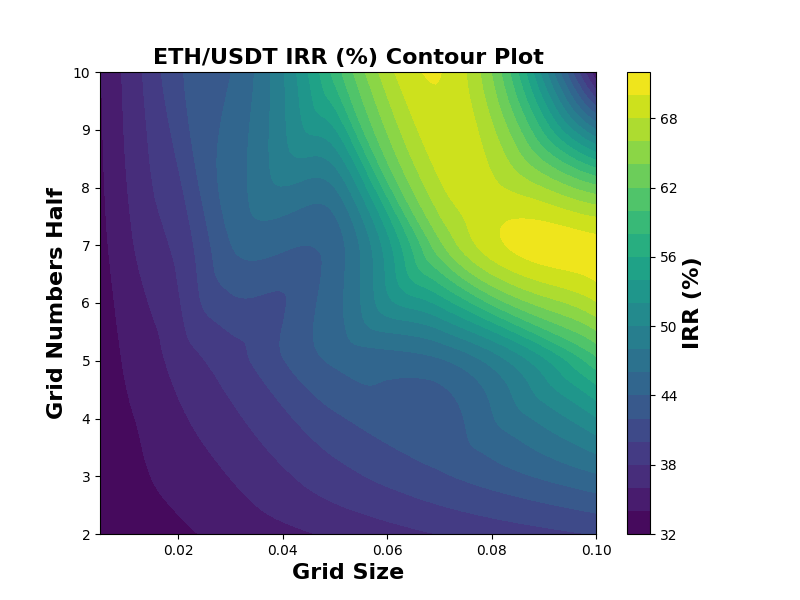}
  \caption{DGT strategy performance of \textit{ETH}}
\end{figure}
\begin{figure}[t]
  \includegraphics[width=\columnwidth]{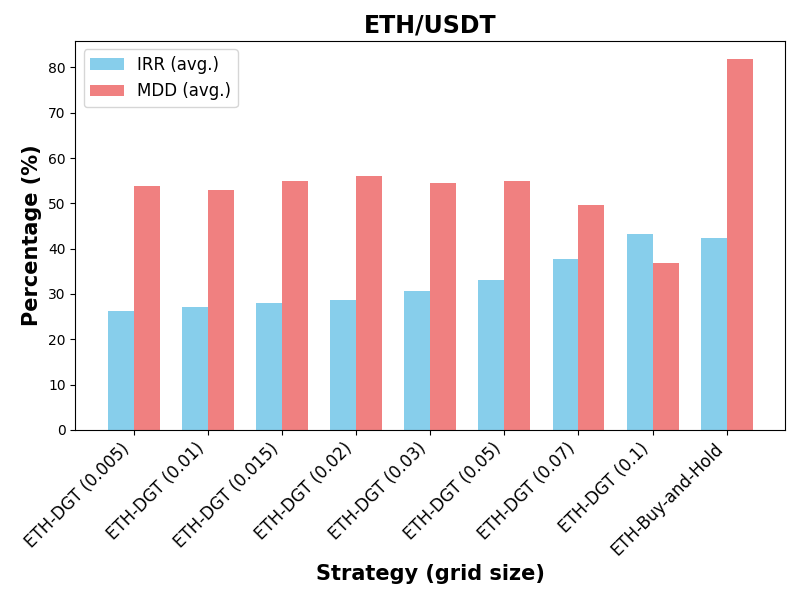}
  \caption{\textit{DGT strategy} performance compared with \textit{buy-and-hold} on ETH}
\end{figure}
\section{Discussion}
\begin{figure*}[h!]
  \includegraphics[width=0.45\linewidth]{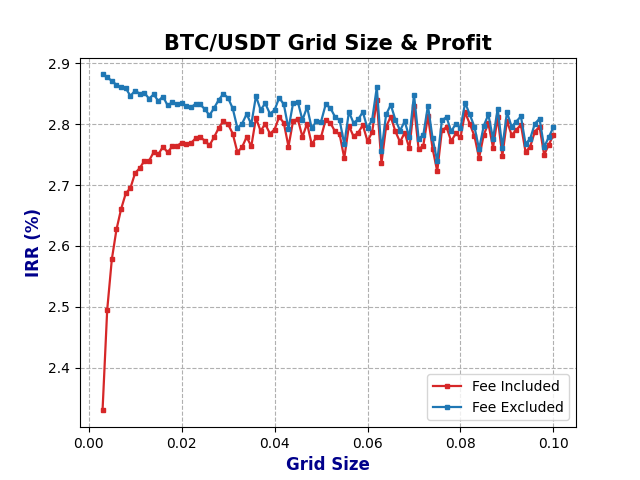} \hfill
  \includegraphics[width=0.45\linewidth]{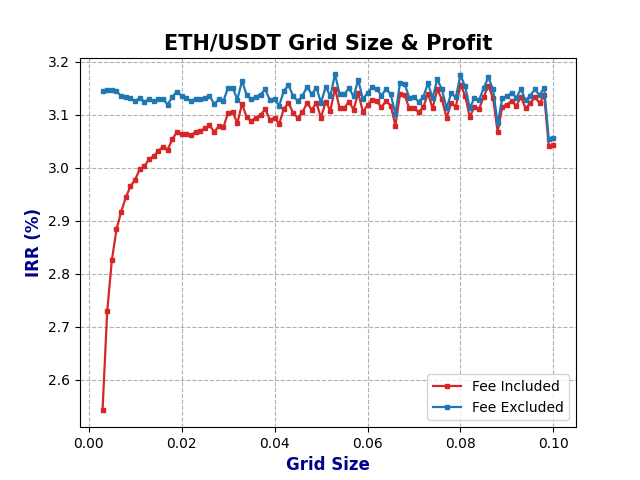}
  \caption {\textit{Traditional grid trading strategy} on BTC \& ETH}
\end{figure*}
Next, we discuss the performance of the \textit{DGT strategy} compared to both the \textit{buy-and-hold strategy} and the \textit{traditional grid trading strategy}.

\subsection{Comparison with \textit{Buy-and-Hold}}

We evaluated two key metrics: \textit{IRR (Internal Rate of Return)} and \textit{MDD (Maximum Drawdown)}. By testing how different \textbf{\textit{DGT}} parameters affect these metrics and comparing the results to the \textit{buy-and-hold strategy}, \textit{figure 5 and 6} present the outcomes of this comparison.

For \textit{BTC}, we observe that the \textit{IRR} of the \textbf{\textit{DGT}} strategy is higher than that of the \textit{buy-and-hold strategy}, while the \textit{MDD} is significantly lower. This shows that the \textbf{\textit{DGT}} strategy not only outperforms the \textit{buy-and-hold strategy} in terms of returns but also manages risk more effectively. Even during market downturns, the strategy ensures relative stability of assets.

In the case of \textit{ETH}, although the \textit{IRR} of the \textbf{\textit{DGT}} strategy is not substantially higher than that of the \textit{buy-and-hold strategy}—likely due to \textit{ETH}’s stronger upward trend during the backtesting period—the \textit{MDD} is still much better controlled. When the market experienced a maximum decline of around 80\%, the \textit{DGT strategy} limited the drawdown to approximately 50\%.

\subsection{Comparison with \textit{Grid Trading Strategy}}
We also compare our \textbf{\textit{DGT}} strategy with the traditional grid trading strategy, demonstrating how the adjustments we made improve overall performance. \textit{Figure 7} shows the \textit{IRR} for a grid strategy where the upper and lower limits perfectly contain the cryptocurrency price from \textit{2021 to July 2024}. For \textit{BTC}, the upper limit is \textit{$80,000$}, and the lower limit is \textit{$10,000$}. For \textit{ETH}, the upper limit is \textit{$5,000$}, and the lower limit is \textit{$500$}.

Several conclusions can be drawn from \textit{Figure 7}. First, transaction fees significantly impact the overall trading results. Without considering fees (represented by the blue line), the \textit{IRR} fluctuates at a consistent level, aligning with our earlier calculation that the expected value of the grid strategy is zero. Next, after considering fees (represented by the red line), as with the \textbf{\textit{DGT}} strategy results, we observe that a smaller \textit{grid size} experiences lower volatility but is more susceptible to transaction fees, while a larger \textit{grid size} exhibits higher volatility due to greater differences in arbitrage levels. Finally, it’s clear that the \textbf{\textit{DGT}} strategy outperforms the \textit{traditional grid trading strategy}, even when applied to a grid with predefined cryptocurrency price boundaries. This is expected, as the \textbf{\textit{DGT}} strategy allows for the reinvestment of profits and more efficient use of capital. Interestingly, we found that many users employ a strategy known as the \textit{"sky and land"} grid strategy on exchanges, which is similar to the \textit{traditional grid trading strategy} discussed here. 

\section{Conclusions and Future Work}
Grid trading has become a widely used trading technique in recent years. However, our research reveals that without any insight into market trends, the expected value of grid trading is effectively zero. This means that investors face a high risk of losing money after taking transaction fees into consideration. To address this, we developed a modified version of the grid trading strategy, known as the \textit{\textbf{DGT}} strategy. While it may seem like a blend of the \textit{buy-and-hold strategy} and the \textit{traditional grid trading strategy}, the \textit{\textbf{DGT}} strategy has significantly outperformed both from \textit{2021 to July 2024}. We hope that our research provides valuable insights for those considering grid trading strategies in the cryptocurrency market.

In this paper, we demonstrated through backtesting that the \textit{\textbf{DGT}} strategy outperforms both the buy and hold strategy and the grid trading strategy. In the future, we will attempt to use mathematical theory to derive the probability of profit and the expected returns of the \textit{\textbf{DGT}} strategy.

\nocite{feasibility2022gridtrading,technical2018cryptocurrency,xu2019machine,2019cryptocurrency}

\section*{Acknowledgments}
We would like to extend our heartfelt gratitude to \textit{Yu-Chen Hung} from the \textit{Department of Economics at National Taiwan University (NTU)} and \textit{Po-Chung Hsieh} from the \textit{Department of Electrical Engineering at NTU }for their valuable insights and thoughtful discussions. Their contributions have been a great source of inspiration and have significantly influenced the successful outcomes of our research.

\bibliography{custom}

\begin{thebibliography}{4}
\providecommand{\natexlab}[1]{#1}

\bibitem[{Francesco~Rundo and Battiato(2019)}]{2019cryptocurrency}
Agatino Luigi di~Stallo Francesco~Rundo, Francesca~Trenta and Sebastiano Battiato. 2019.
\newblock Grid trading system robot (gtsbot): A novel mathematical algorithm for trading fx market.
\newblock \emph{Applied Sciences}, 9(9):1796.

\bibitem[{Jia(2022)}]{feasibility2022gridtrading}
Ruixin Jia. 2022.
\newblock \href {https://doi.org/10.4108/eai.18-11-2022.2327164} {The feasibility of grid trading approach for bitcoin based on backtesting}.
\newblock In \emph{Proceedings of the 4th International Conference on Economic Management and Model Engineering, ICEMME 2022, November 18-20, 2022, Nanjing, China}.

\bibitem[{Kamrat et~al.(2018)Kamrat, Suesangiamsakul, and Marukatat}]{technical2018cryptocurrency}
Sirapop Kamrat, Napasorn Suesangiamsakul, and Rangsipan Marukatat. 2018.
\newblock Technical analysis for cryptocurrency trading on mobile phones.
\newblock In \emph{The 2018 Technology Innovation Management and Engineering Science International Conference (TIMES-iCON2018)}, pages 1--4.

\bibitem[{Xu et~al.(2019)Xu, Hu, Yang, Zhang, and Ye}]{xu2019machine}
Jia Xu, Kun Hu, Guang Yang, Jian Zhang, and Jianxing Ye. 2019.
\newblock Using machine learning for cryptocurrency trading.
\newblock In \emph{2019 IEEE International Conference on Industrial Cyber Physical Systems (ICPS)}, pages 647--652.

\end{thebibliography}

\appendix
\section{Code Availability}

The code used in this paper is available on GitHub at the following link: \href{https://github.com/colachenkc/Dynamic-Grid-Trading}{Source codes}.

\end{document}